\documentclass[11pt]{article}

\usepackage[T1]{fontenc}
\usepackage[utf8]{inputenc}
\usepackage{lmodern}
\usepackage{microtype}

\usepackage[margin=1in]{geometry}

\usepackage{graphicx}
\graphicspath{{figures/}}

\usepackage{booktabs}
\usepackage{array}
\usepackage{multirow}

\usepackage[font=small,labelfont=bf,labelsep=colon]{caption}

\usepackage{authblk}

\usepackage[hidelinks,breaklinks=true]{hyperref}

\usepackage[super,numbers,sort&compress]{natbib}
\bibpunct{}{}{,}{s}{}{,}

\title{A strongly annotated passive acoustic dataset for tropical bird monitoring}

\author[1,5]{Daniela Ruiz}
\author[2]{Juan Sebasti\'an Ulloa}
\author[1]{Zhongqi Miao}
\author[2]{Nicol\'as Betancourt}
\author[2]{Maria Paula Toro-G\'omez}
\author[1,5]{Andr\'es Hern\'andez}
\author[1]{Bruno Demuro}
\author[2]{Eliana Barona-Cort\'es}
\author[3]{Angela M.\ Mendoza-Henao}
\author[2]{Andr\'es Sierra-Ricaurte}
\author[4]{Sebastian P\'erez-Pe\~na}
\author[1]{Rahul Dodhia}
\author[5]{Pablo Arbel\'aez}
\author[1]{Juan Lavista}

\affil[1]{Microsoft AI for Good Research Lab, Redmond, Washington, United States}
\affil[2]{Instituto de Investigaci\'on de Recursos Biol\'ogicos Alexander von Humboldt, Bogot\'a, Colombia}
\affil[3]{Fundaci\'on Manacus, Red Ecoac\'ustica Colombiana, Cali, Colombia}
\affil[4]{Louisiana State University, Baton Rouge, United States, Museum of Natural Sciences}
\affil[5]{Universidad de Los Andes, Bogot\'a, Colombia, Center for Research and Formation in Artificial Intelligence}

\date{}

\begin{document}
\maketitle

\begin{abstract}
Passive acoustic monitoring enables continuous, non-invasive biodiversity assessment across diverse ecosystems. The scale of these datasets has driven the adoption of machine learning, with supervised approaches showing strong performance. However, supervised methods require time-resolved annotated datasets, which remain scarce, especially in complex tropical soundscapes. We present PteroSet, a curated dataset of strongly annotated Neotropical bird vocalizations recorded in Puerto Asís (Putumayo) and Pivijay (Magdalena), Colombia, between 2023 and 2025. The dataset comprises 563 recordings (73.62 h) and 15,372 time--frequency annotations, including 6,702 events identified to the species level across 168 species. We release the annotations in a COCO-inspired JSON schema that unifies audio files, taxonomic categories, and labels for machine learning workflows. Beyond providing annotated data, PteroSet serves as a realistic benchmark that highlights key characteristics of tropical soundscapes, including acoustic co-occurrence and domain shift across recording sites. We provide a deep learning baseline for binary bird detection, demonstrating PteroSet's usability and the challenges it presents.

\end{abstract}

\section*{Background \& Summary}
Passive acoustic monitoring (PAM) has emerged as a powerful, non-invasive approach for studying biodiversity across broad spatial and temporal scales. Researchers deploy autonomous recording units to enable continuous monitoring of vocal species without direct observation or capture\cite{ref1,ref2}. This approach is particularly valuable in remote or dense habitats, where traditional survey methods are invasive, costly, logistically challenging, or prone to observer bias\cite{ref3}. Advances in recording technology and reduced storage costs enable PAM programs to generate large volumes of audio data. These datasets create new opportunities for ecological research and conservation, but they also pose significant analytical challenges\cite{ref4,ref5}. The scale of modern PAM datasets makes manual inspection and annotation impractical, driving the growing adoption of machine learning (ML) methods for automated sound detection and classification\cite{ref6}. Supervised ML approaches, particularly deep learning models, have demonstrated strong performance in recognizing complex acoustic patterns\cite{ref4,ref7,ref8,ref9,ref10,ref11}. Nonetheless, their success depends on large, high-quality annotated datasets that support reproducibility, benchmarking, and generalization across diverse taxa and regions.

Birds are among the most important and extensively studied taxonomic groups in bioacoustics. As highly vocal organisms with well-characterized acoustic repertoires, birds are widely used as indicators of ecosystem health, habitat change, and biodiversity trends\cite{ref7,ref12,ref13}. Consequently, automated bird detection has become a central task in bioacoustics and conservation-oriented machine learning research.\cite{ref6} A growing body of work has introduced public datasets and benchmarks\cite{ref14,ref15,ref16,ref17,ref18,ref19,ref20,ref21,ref22,ref23,ref24,ref25} for bird sound analysis across diverse ecological settings, recording conditions, and learning tasks, alongside international challenges such as BirdCLEF\cite{ref26} and DCASE\cite{ref27}. In parallel, community-driven repositories such as Xeno-canto\cite{ref28} have made large-scale bird recordings openly available, though they are typically weakly annotated and limited to clip-level labels, which introduce recording biases. These resources have enabled the development of bioacoustics foundation models, such as BirdNET\cite{ref29} and Perch\cite{ref30}, demonstrating the feasibility of automated bird monitoring at scale. However, most existing bird datasets come from temperate regions, and models trained on them often underperform in tropical landscapes.

Tropical regions, despite their exceptional biodiversity, remain underrepresented in global data infrastructures such as the Global Biodiversity Information Facility (GBIF), where data publication and occurrence records are concentrated in Europe and North America\cite{ref31}. This imbalance is even more pronounced in the Neotropics, particularly in countries such as Colombia, widely recognized as the most bird-rich country globally\cite{ref32}. Only a few datasets address this gap. These include the Colombia--Costa Rica coffee-farm dataset, which provides 34 hours of recordings from 89 species in Andean mid-elevation ecosystems\cite{ref33}, and the Peruvian lowland Amazon dataset, which includes 21 hours and 132 species\cite{ref34}. The multi-country WABAD compilation\cite{ref35} further expands coverage and includes a Colombian subset with 6.7 hours of recordings and 165 species across Andean coffee landscapes and the Orinoco llanos.

To address the mismatch between biological richness and digital representation in bioacoustics, we present PteroSet, a curated dataset of bird vocalizations with robust manual annotations, recorded in Neotropical environments in Colombia. We collected recordings from two ecologically contrasting, undersampled landscapes: the historically transformed Magdalena Medio and the Amazonian foothills of Putumayo, a region experiencing accelerating habitat loss driven by hydrocarbon extraction and agricultural expansion. Together, these sites capture acoustically complex tropical soundscapes and provide a baseline for avian communities in regions undergoing rapid environmental transformation.

More specifically, the main contributions of this work are as follows:

\begin{enumerate}
\def\labelenumi{\arabic{enumi}.}
\item
  PteroSet dataset. We provide 563 time-lapse recordings (73.62 hours) and 15,372 time--frequency event annotations, including 6,702 species-level labels for 168 Neotropical bird species. This dataset substantially expands the scale of highly annotated audio from Colombia and covers previously unsampled biogeographic zones.
\item
  Realistic tropical benchmark. PteroSet highlights challenges typical of Neotropical soundscapes, such as dense acoustic overlap and variability across recording sites, which are often underrepresented in temperate datasets.
\item
  Annotation schema. We introduce a practical JSON structure adapted from the Common Objects in Context (COCO)\cite{ref36} image annotation format to meet the specific requirements of bioacoustic data and to improve dataset usability.
\item
  Reference baseline. We train a ResNet-18 binary bird-presence detector to provide a reproducible baseline for evaluating models on PteroSet.
\item
  Open-access resources. We release the PteroSet dataset and associated code as open-source resources to support reproducibility and to facilitate future research.
\end{enumerate}

We summarize the complete PteroSet workflow, from data collection to machine learning validation, in Fig. 1.

\begin{figure}[!htbp]
  \centering
  \includegraphics[width=\linewidth]{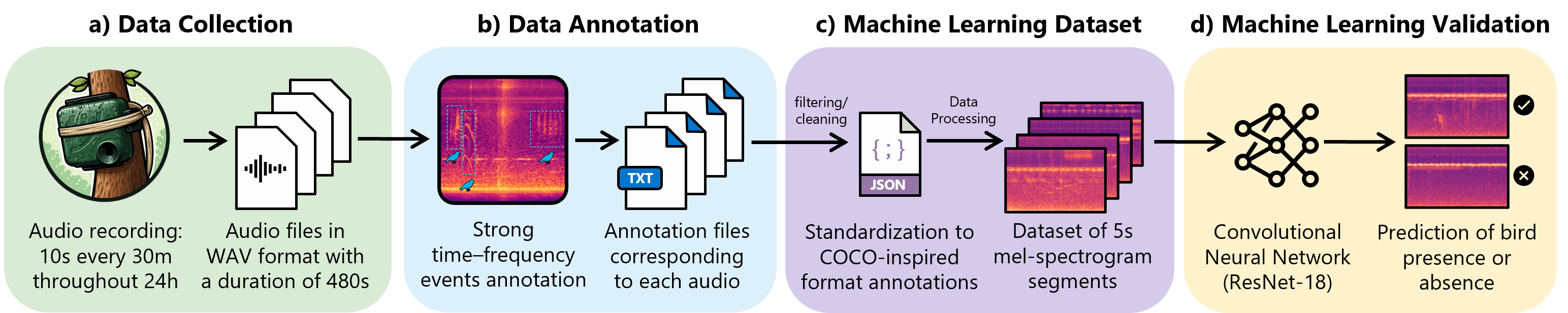}
  \caption{Overview of the PteroSet data pipeline and baseline modeling approach. (a) Data Collection: We collect audio data using a time-lapse protocol of 10-second recordings every 30 minutes over 24 hours, yielding 480 seconds of WAV files per day. (b) Data Annotation: We manually annotate each recording in Raven as strong time--frequency events (t\_min, t\_max, f\_min, f\_max) and store them as a plain-text file per audio recording. (c) Machine Learning Dataset: We filter the data to retain only bird annotations and recordings, then standardize it into a COCO-inspired JSON format. Using these annotations, we generate 5-second mel-spectrogram segments to produce inputs for the machine learning model. (d) Machine Learning Validation: We train a ResNet-18 convolutional network to predict bird presence or absence for each 5-second segment, serving as a reference baseline for the dataset.}
  \label{fig:overview}
\end{figure}

\section*{Methods}
\subsection*{Data Collection}

We monitor vocal activity in Neotropical bird communities at two study sites in Colombia: Puerto Asís (Putumayo Department) and Pivijay (Magdalena Department). Both sites are part of environmental characterization and monitoring programs implemented to comply with regulatory requirements for hydrocarbon exploration and production. These regulatory contexts, while reflecting significant anthropogenic pressure, provide a rare opportunity to systematically document avian acoustic diversity in biologically rich yet chronically undersampled landscapes in global biodiversity repositories. The study areas represent contrasting ecological contexts within the Andean--Amazonian transition zone (Putumayo) and the Caribbean lowlands of northern Colombia (Magdalena). Putumayo sits at the interface of Andean and Amazonian biotas, a region of exceptional species turnover and endemism that is undergoing rapid landscape transformation. Magdalena, by contrast, represents a more historically intervened matrix where remnant avian assemblages persist within a highly fragmented agricultural and pastoral landscape. Both regions are characterized by high avian diversity and heterogeneous gradients of anthropogenic pressure. Together, they constitute a gradient of transformation histories, with one capturing a landscape in early-stage conversion and the other in advanced degradation. Recording sites and the spatial distribution of deployed autonomous recorders are shown in Fig. 2.

We conduct PAM using autonomous acoustic sensors (AudioMoth\cite{ref37}) equipped with internal omnidirectional microphones. At each site, we mount recorders on trees or wooden supports at approximately 1.5\,m above ground level. We program the devices to record 1\,min every 30\,min, continuously over a 24\,h period, yielding 48\,min of audio per device per day (sampling rate: 192 kHz; 16-bit resolution). We collect audio recordings during five monitoring seasons spanning 2023 to 2025 (Table 1) and store them in uncompressed WAV format for subsequent annotation and analysis. Although the broader monitoring framework includes multiple taxonomic groups, we restrict the present study to bird vocalizations and exclude acoustic data from other taxa and non-biological sound sources from the analysis. From the full dataset, we select days with low levels of abiotic noise for expert annotation to ensure high bird detectability.

\begin{table}[!htbp]
  \centering
  \caption{Monitoring study sites, seasons, project names, and recording periods for passive acoustic data collection conducted in Colombia between 2023 and 2025.}
  \label{tab:sites}
  \small
  \begin{tabular}{lccll}
    \toprule
    \textbf{Site} & \textbf{Season} & \textbf{Project Name} & \textbf{Date start} & \textbf{Date end} \\
                  &                 &                       & \textbf{(yyyy-mm-dd)} & \textbf{(yyyy-mm-dd)} \\
    \midrule
    Magdalena -- Pivijay       & 1 & MAP1 & 2023-04-11 & 2023-06-07 \\
    Putumayo -- Puerto As\'is  & 2 & PPA1 & 2024-02-07 & 2024-03-03 \\
    Putumayo -- Puerto As\'is  & 3 & PPA2 & 2024-12-14 & 2024-12-28 \\
    Putumayo -- Puerto As\'is  & 4 & PPA3 & 2025-02-27 & 2025-03-25 \\
    Putumayo -- Puerto As\'is  & 5 & PPA4 & 2025-06-19 & 2025-07-02 \\
    \bottomrule
  \end{tabular}
\end{table}

\begin{figure}[!htbp]
  \centering
  \includegraphics[width=\linewidth]{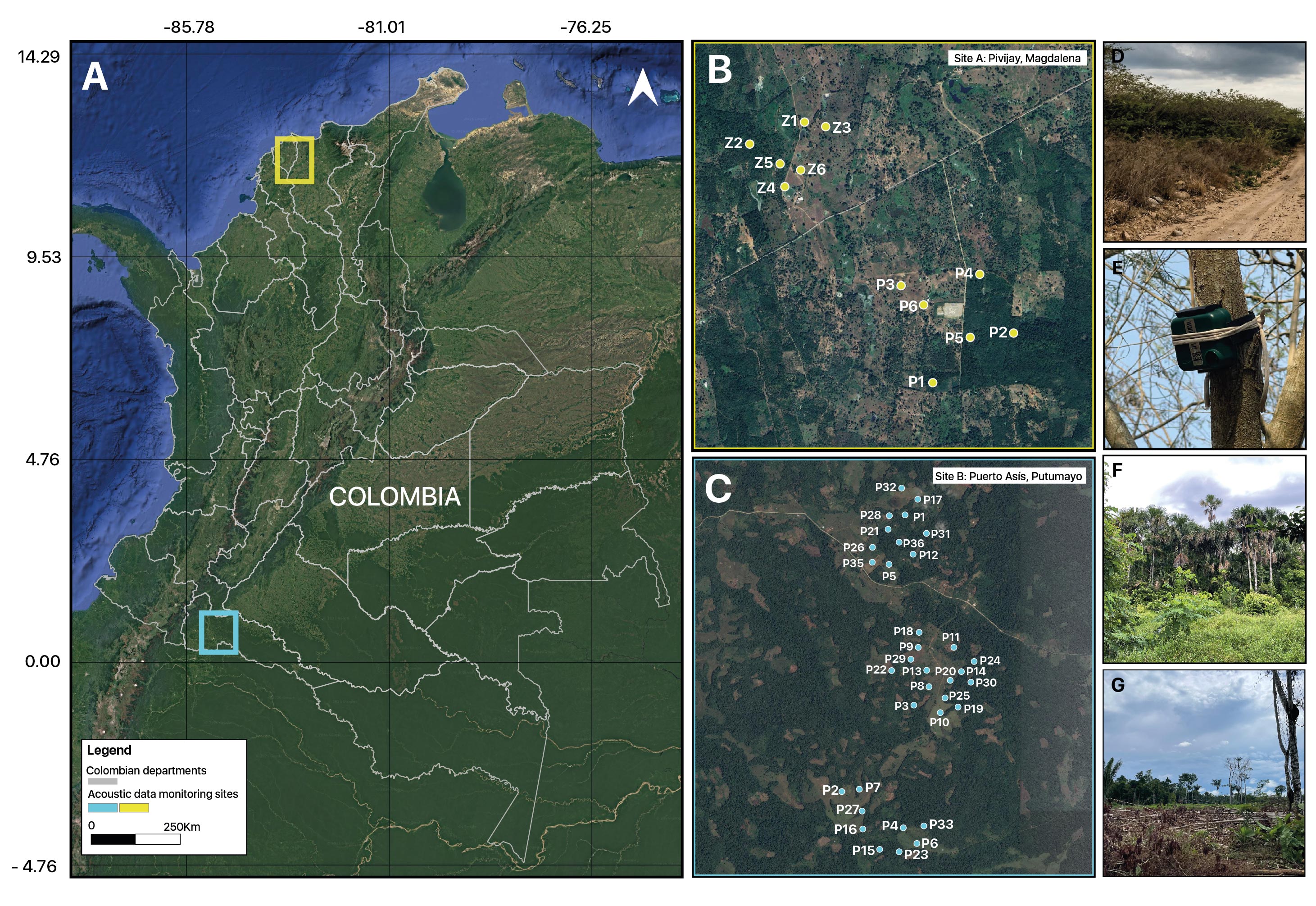}
  \caption{Overview of recording locations. (a) Map of Colombia showing the two study regions where we conduct passive acoustic monitoring deployments: Pivijay (Magdalena) and Puerto Asís (Putumayo). (b--c) Detailed maps showing the specific locations of autonomous recording device deployments at each study site. (d--g) Representative in situ photographs illustrating the deployment environments and examples of recording devices installed in the field.}
  \label{fig:map}
\end{figure}

\subsection*{Data Annotation}

We develop a standardized annotation protocol to generate high-quality, time--frequency-bounded labels for training and evaluating bird detection and identification models. All annotations correspond to individual vocal events and include explicit onset, offset, and frequency bounds.

To facilitate expert inspection of long-duration recordings, we convert audio files into time-lapse spectrogram representations that summarize acoustic activity over a full 24-hour period. We generate time-lapse files by extracting the first 10 seconds from each of the 48 recordings collected per day and concatenating them into a single 480-second file that represents acoustic samples from different times of day (Fig. 3). We adopt this sub-sampling protocol to preserve temporal coverage across the full 24-hour cycle while keeping the manual annotation workload tractable. We carefully select days for time-lapse generation to avoid periods of intense rainfall, thereby maximizing the detectability of bird vocal activity.

Four specialists (AFSR, EBC, MPTG, SPP) with prior experience identifying Neotropical bird vocalizations perform all annotations through detailed visual and auditory inspections of spectrograms using Raven Pro software\cite{ref38}. Annotators manually identify and delimit individual bird vocal events and assign species identities based on diagnostic acoustic features. Experts annotate only acoustic signals that are clearly identified both acoustically and visually in the spectrogram, systematically excluding faint or ambiguous vocalizations.

\begin{figure}[!htbp]
  \centering
  \includegraphics[width=\linewidth]{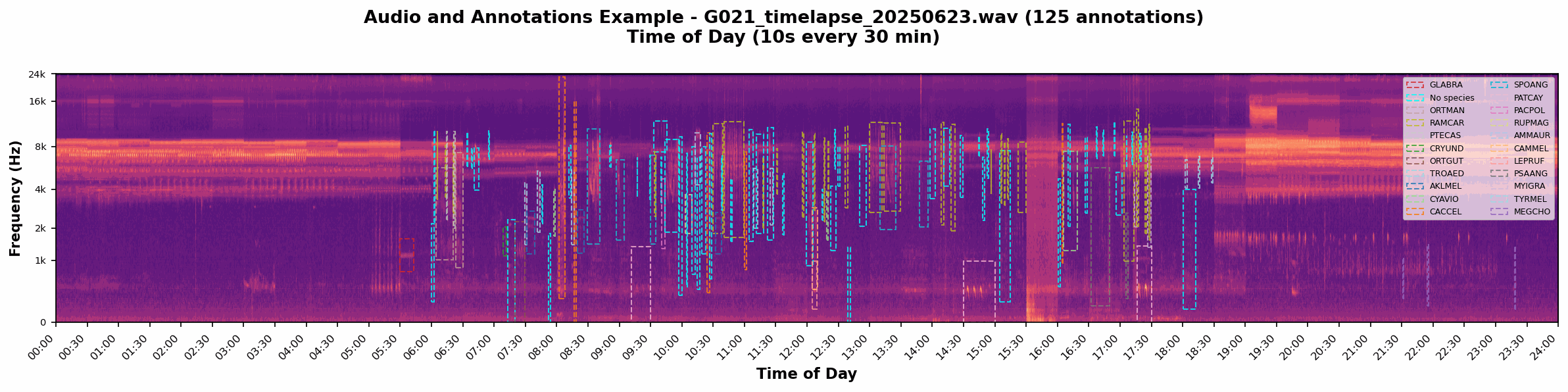}
  \caption{Time-lapse spectrogram of a 24-hour acoustic recording with annotations. We create the audio by concatenating the first 10 seconds of each of the 48 recordings collected at 30-minute intervals throughout the day. Frequency (Hz) is shown on the y-axis and time of day on the x-axis. Dashed boxes indicate manually annotated bird vocalizations (125 annotations from 22 different species in this example).}
  \label{fig:timelapse}
\end{figure}

We label species using a standardized alphanumeric coding system that uses the first three letters of the genus followed by the species epithet (e.g., \emph{Megascops choliba} is coded as MEGCHO). We do not assign explicit signal-quality categories during annotation because we focus on obtaining precise temporal labels for supervised learning and model evaluation.

\subsection*{Computational processing}

Because we generate all annotations manually, we perform several consistency checks before downstream processing. First, we verify that all column names across annotation files are spelled, capitalized, and ordered identically, and we correct minor typographical errors when detected. Second, we standardize categorical values in the Tipo, ID, and Determination columns to enforce a consistent vocabulary across the entire dataset. This step is critical to avoid silent label mismatches during automated parsing and filtering. Third, we verify a one-to-one correspondence between audio recordings and annotation files, ensuring that every WAV file has an associated TXT annotation and vice versa.

We resample all audio recordings from 192 kHz to 48 kHz for analysis, while we release the original 192 kHz recordings to ensure full reproducibility. We use the original sampling rate for bat monitoring during field deployment. However, the present dataset focuses on bird vocalizations, which occur well below the ultrasonic range. As a result, frequencies above 24 kHz do not contain biologically relevant information for the target taxa but substantially increase storage requirements and computational cost during processing and analysis. We inspect the annotations and confirm that only 130 of 15,372 annotated events have a maximum frequency exceeding 24 kHz, and these cases correspond to recording equipment artifacts rather than bird signals. Therefore, we recommend resampling the data to 48 kHz for standard use and capping the maximum frequency values in the annotation tables at 24 kHz to ensure consistency with the processed audio representation.

Beyond ensuring internal consistency, we design these cleaning steps to facilitate future reuse and integration with other bioacoustics resources. Differences in annotation schemas, labeling practices, and metadata structures hinder the combination of recordings and limit large-scale machine learning and comparative analyses. While no single standard accommodates all use cases, we adopt a flexible representation that captures core elements and reduces these barriers. To this end, we develop a unified annotation schema inspired by the COCO\cite{ref36} format commonly used in computer vision and adapt it to audio data by replacing image-based fields with sound-specific metadata such as audio duration, sampling rate, and file paths. We design the format to support both strong annotations (with explicit time--frequency boundaries) and weak annotations (with partial or missing time--frequency information), enabling integration of datasets generated by different annotation tools and research workflows. We convert all annotations into the unified JSON structure to ensure consistent preprocessing across the combined dataset.

\section*{Data Records}
PteroSet comprises 563 audio recordings from five projects (MAP1, PPA1, PPA2, PPA3, and PPA4), totaling 73.62 hours of audio. The recordings are relatively uniform in length. For MAP1 and PPA2-4, most recordings have the expected duration of 480 seconds (48 segments x 10 s), although a few PPA4 files are shorter (440, 460, or 470 seconds) due to missing segments caused by recorder malfunctions. PPA1 recordings are 433 seconds long because we applied a 1-second crossfade between consecutive 10-second segments during timelapse compilation.

PteroSet contains 15372 class-level and 6702 species-level annotations. Annotation segments are typically short, with a median duration of 0.818 s, a mean of 1.697 s, and a standard deviation of 2.140 s. In the frequency domain, annotations span a low-frequency onset of 1,702.6 $\pm$ 1,246.9 Hz (median 1,411.8 Hz) and a high-frequency offset of 5,192.7 $\pm$ 3,344.4 Hz (median 4,743.5 Hz), yielding a mean bandwidth of 3,490.0 $\pm$ 3,077.0 Hz (median 2,710.6 Hz). Most vocalizations fall within the 1--5 kHz band characteristic of Neotropical bird assemblages.

We generate the original annotations with the Raven Pro\cite{ref38} sound analysis software and store them as plain-text tabular files, one per time-lapse audio recording. We describe the annotation format and all associated fields in Table 2. Each annotation file contains temporal and spectral information for individual sound events, including start and end times and lower and upper frequency bounds. In addition to acoustic parameters, each annotation includes three categorical fields: Tipo, ID, and Determination. The Tipo field denotes the event\textquotesingle s broad soundscape category. Because this study focuses exclusively on biological sounds, we retain only BIO-labeled annotations for further processing. The ID field specifies the taxonomic group responsible for each sound event. We exclude all non-Avian annotations and retain only events classified as AVEVOC in the final dataset. Finally, the Determination field encodes species-level identification using standardized alphanumeric species codes. We provide a complete list of valid species codes and their corresponding taxa as a separate CSV file distributed with the dataset.

\begin{table}[!htbp]
  \centering
  \caption{Description of fields in the original Raven Pro annotation format.}
  \label{tab:raven-fields}
  \small
  \begin{tabular}{p{0.18\linewidth}p{0.76\linewidth}}
    \toprule
    \textbf{Field} & \textbf{Description} \\
    \midrule
    Selection       & A unique number that identifies each individual annotation within the file and preserves its row order. \\
    View            & Indicates the visual display (such as spectrogram or waveform) in which the annotation was created. \\
    Channel         & Specifies the audio channel from which the annotated sound originates, especially relevant for stereo recordings. \\
    Begin Time (s)  & The time is in seconds from the start of the audio file at which the annotated sound begins. \\
    End Time (s)    & The time in seconds from the start of the audio file at which the annotated sound ends. \\
    Low Freq (Hz)   & The lowest frequency of the annotated sound in Hertz represents the lower bound of the sound energy. \\
    High Freq (Hz)  & The highest frequency of the annotated sound in Hertz represents the upper bound of the sound energy. \\
    Tipo            & The general soundscape category. \\
    ID              & The taxonomic group causing the sound event. \\
    Determination   & The species identity using standardized alphanumeric species codes. \\
    \bottomrule
  \end{tabular}
\end{table}

We integrate all annotation files into the structured JSON format described in Table 3, applying the data cleaning and filtering procedures outlined above to retain only bird-related annotations. We store dataset-level information in the info object, which records descriptive metadata, licensing, versioning, and authorship. We define biological labels in the categories array, assigning each category a unique identifier and a taxonomic name. We describe audio recordings in the sounds array, including file paths, duration, sampling rate, and geographic coordinates of the recording location. We represent individual sound events in the annotations array and explicitly link them to both sound files and categories via unique identifiers. Each annotation specifies temporal boundaries (t\_min, t\_max) and spectral boundaries (f\_min, f\_max), enabling precise localization of signals in the time--frequency domain.

\begin{table}[!htbp]
  \centering
  \caption{Specification of the bioacoustic annotation JSON format, including dataset-level metadata, category definitions, sound file information, and time--frequency annotations.}
  \label{tab:json-schema}
  \small
  \begin{tabular}{p{0.14\linewidth}p{0.20\linewidth}p{0.58\linewidth}}
    \toprule
    \textbf{Section} & \textbf{Field} & \textbf{Description} \\
    \midrule
    \multirow{7}{*}{\textbf{info}}
        & title             & Title of the bioacoustic dataset. \\
        & license           & License under which the dataset is released. \\
        & publication\_date & Date on which the dataset was published. \\
        & description       & Brief textual summary describing the dataset contents. \\
        & creators          & List of dataset creators, including name and affiliation. \\
        & version           & Dataset version number. \\
        & url               & Persistent URL or landing page for the dataset. \\
    \midrule
    \multirow{2}{*}{\textbf{categories}}
        & id                & Unique identifier for each category. \\
        & name              & Scientific or common name of the annotated category or species. \\
    \midrule
    \multirow{7}{*}{\textbf{sounds}}
        & id                & Unique identifier for each sound file. \\
        & file\_name\_path  & Absolute path to the audio file. \\
        & duration          & Total duration of the audio file in seconds. \\
        & sample\_rate      & Audio sampling rate in Hertz (Hz). \\
        & latitude          & Latitude of the recording location in decimal degrees. \\
        & longitude         & Longitude of the recording location in decimal degrees. \\
        & date\_recorded    & Date on which the recording was made. \\
    \midrule
    \multirow{8}{*}{\textbf{annotations}}
        & anno\_id          & Unique identifier for each annotation. \\
        & sound\_id         & Identifier linking the annotation to a sound file. \\
        & category\_id      & Identifier linking the annotation to a category. \\
        & category          & Name of the annotated category or species. \\
        & t\_min            & Start time of the annotation in seconds from the beginning of the audio file. \\
        & t\_max            & End time of the annotation. Seconds from the beginning of the audio file. \\
        & f\_min            & Minimum frequency of the annotation in Hertz (Hz). \\
        & f\_max            & Maximum frequency of the annotation in Hertz (Hz). \\
    \bottomrule
  \end{tabular}
\end{table}

We make the complete dataset publicly available on Zenodo\cite{ref39}, including raw annotation files and processed JSON records. We release all data under the Creative Commons Attribution 4.0 International license, ensuring unrestricted access and reuse.

\subsection*{Data Overview}

\begin{figure}[!htbp]
  \centering
  \includegraphics[width=\linewidth]{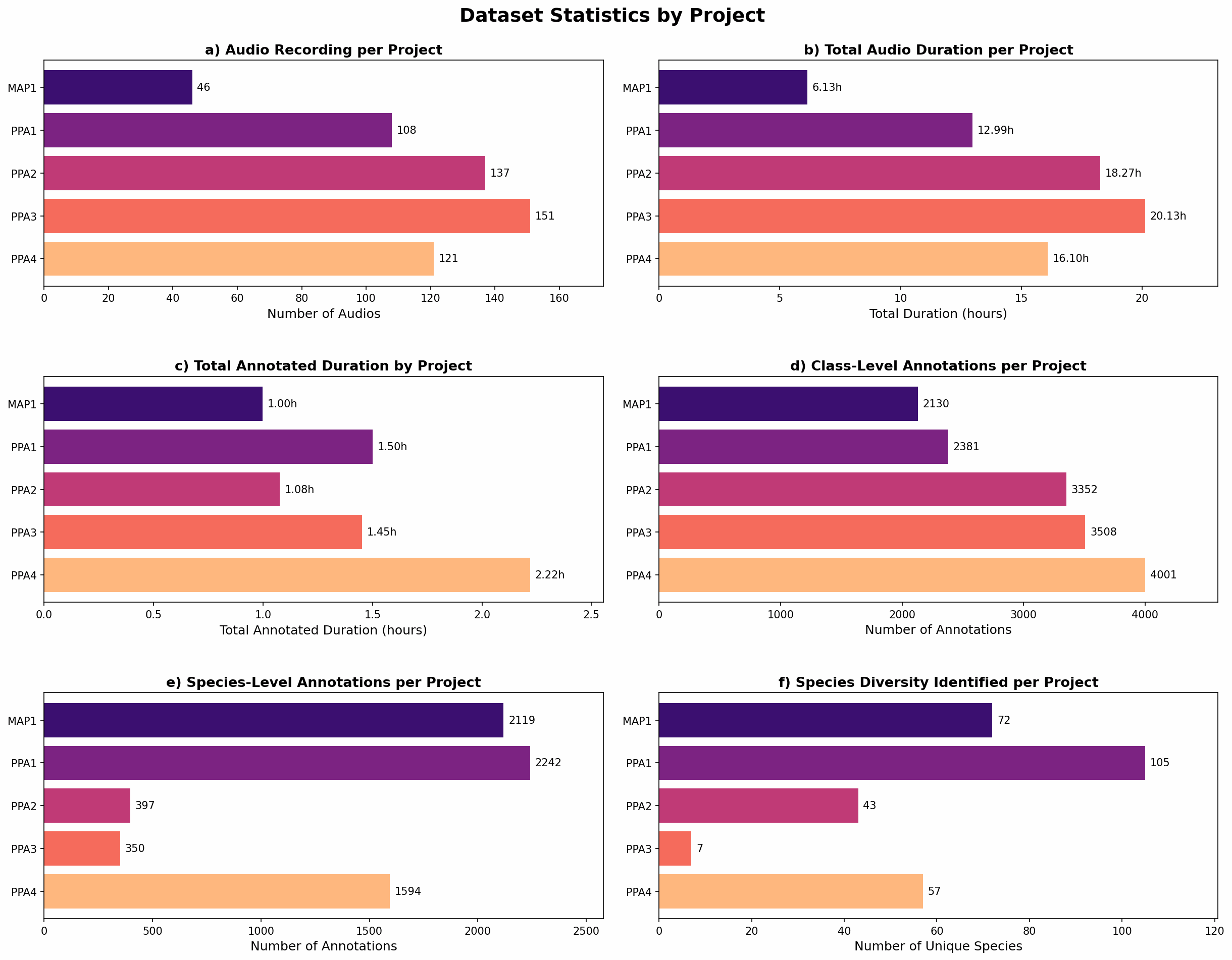}
  \caption{Summary statistics by project. (a) Number of audio recordings per project. (b) Total audio duration (hours) per project. (c) Total annotated duration (hours) per project. (d) Number of class-level annotations per project. (e) Number of species-level annotations per project. (f) Number of unique species identified per project.}
  \label{fig:stats}
\end{figure}

Fig. 4 summarizes the distribution of recordings, durations, and annotations across projects. The number of recordings per project ranges from 46 in MAP1 to 151 in PPA3 (Fig. 4a). PPA3 has the largest total audio duration (20.13 h), whereas MAP1 has the smallest (6.13 h) (Fig. 4b). Fig. 4c shows the total annotated duration per project, indicating that approximately 6.66 hours ($\sim$9\%) of the 73.62 hours of recordings contain bird vocalizations. Annotation density varies considerably between projects. The number of class-level annotations per project (Fig. 4d) ranges from 2,130 in MAP1 to 4,001 in PPA4. PPA4, PPA3, and PPA2 contain the highest numbers of class-level annotations but the lowest numbers of species-level annotations (Fig. 4e). In contrast, PPA1 and MAP1 show a high proportion of species-level annotations relative to class-level annotations, indicating that annotators identify most vocalizations at the species level in these projects. Species diversity also differs substantially across projects (Fig. 4f). PPA1 contains the highest number of unique species, followed by MAP1 and PPA4. Across all projects combined, the dataset includes 168 unique species. The most frequently annotated species include \emph{Cyanocorax violaceus} (914 annotations), \emph{Psarocolius angustifrons} (535), \emph{Crypturellus cinereus} (418), and \emph{Pitangus sulphuratus} (406). Annotators do not assign all annotations to the species level due to taxonomic uncertainty, so the true species diversity within each project may exceed current estimates. Differences in annotation strategy across projects further contribute to this uncertainty. In PPA1, annotators label all vocalizing bird species at the highest possible taxonomic resolution. In contrast, in PPA2--PPA4, annotators apply a more selective strategy that prioritizes key species of interest, particularly \emph{Cyanocorax violaceus} and \emph{Crypturellus cinereus}. This uneven annotation effort likely drives variation in species diversity and contributes to class imbalance, with the dataset remaining skewed toward a subset of frequently annotated taxa.

\section*{Technical Validation}
We process audio data to generate inputs suitable for deep learning. We transform all recordings into time--frequency representations as spectrograms. Based on an exploratory analysis of annotation durations, we find that most bird vocalizations are under 2 seconds, and less than 8\% exceed 5 seconds. We therefore select a 5-second window length to ensure that most vocalizations are fully captured within a single input sample.

We apply a sliding window with a 1-second step size (corresponding to a 4-second overlap) within each 10-second audio segment used to construct the time-lapse recordings. We exclude windows that cross segment boundaries to avoid mixing acoustic information from different temporal segments, yielding six windows per segment. This approach ensures that each spectrogram corresponds to a single recording segment while increasing the number of training samples through overlapping segmentation. For PPA1 recordings, we adjust segment boundaries to account for the 1-second crossfade between consecutive segments, resulting in a 9-second stride instead of 10.

We convert each 5-second audio window into a mel-spectrogram using a 2048-point fast Fourier transform, a 512-sample hop length, and 224 mel-frequency bands. We compute spectrograms in the power--mel domain and convert them to a decibel scale with an 80 dB dynamic range. All recordings use a 48 kHz sampling rate, yielding a frequency range of 0--24 kHz. We precompute and store all spectrograms before model training to ensure reproducibility and to separate input preprocessing from model optimization.

We formulate the task as a binary detection problem, classifying each spectrogram window as containing bird vocalizations or not. We derive window-level labels from the underlying time--frequency annotations using an overlap criterion (Fig. 5). We label a window as positive if it overlaps with at least one annotation, regardless of whether the annotation is fully contained, partially overlaps the boundaries, spans the entire window, or co-occurs with other annotations. We label windows with no overlapping annotations as negative.

Across the full dataset, 160,244 5-second windows are available for binary detection, of which 20.9\% are positive. Positive rates vary across projects: PPA2 shows the lowest proportion (17.3\%), followed by PPA3 (18.7\%), PPA4 (21.4\%), and PPA1 (22.2\%), while MAP1 exhibits a higher rate (34.5\%). Within each project, positive windows rarely occur in isolation. On average, each positive window overlaps 1.62--1.86 annotations, with a median of 1 and a maximum of 6--10 simultaneous annotations. These values quantify the dense acoustic co-occurrence characteristic of Neotropical soundscapes.

We divide the dataset using a leave-one-project-out cross-validation strategy with five folds, each corresponding to a distinct project. In each fold, we hold out all recordings from one project as the test set and use the remaining four projects for training and validation. Specifically, Fold 0 uses MAP1 as the test set, Fold 1 uses PPA1, Fold 2 uses PPA2, Fold 3 uses PPA3, and Fold 4 uses PPA4. This strategy enforces strict temporal and spatial separation between training and evaluation data, minimizes data leakage, and provides a robust assessment of model generalization to unseen conditions. As described above, we generate training samples using a sliding window with a 1-second step. However, we use only non-overlapping windows for evaluation to prevent inflated performance estimates.

\begin{figure}[!htbp]
  \centering
  \includegraphics[width=\linewidth]{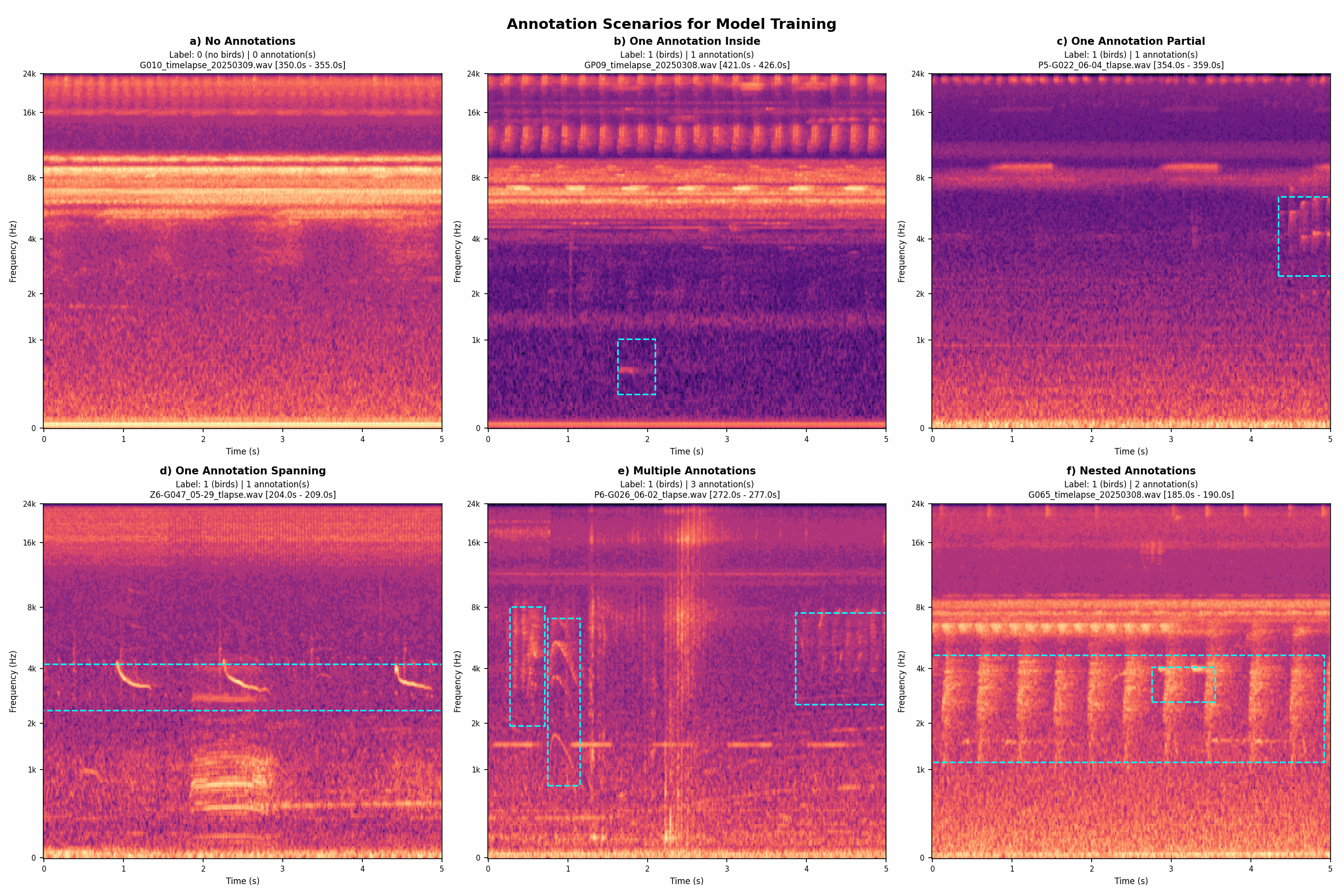}
  \caption{Annotation scenarios for binary classification training. Each panel shows a 5-second mel spectrogram window with overlaid annotations (cyan dashed boxes). A window is labeled positive (birds present) if any annotation overlaps it, regardless of the degree of overlap. (a) No annotations overlap the window, yielding a negative label. (b) One annotation falls entirely within the window. (c) An annotation partially overlaps the window boundary. (d) A single annotation spans the entire window. (e) Multiple annotations overlap the same window. (f) Nested annotations, where one bounding box is fully contained within another. Cases (b)--(f) all yield a positive label.}
  \label{fig:scenarios}
\end{figure}

We train a ResNet-18 backbone initialized with ImageNet-pretrained weights. To adapt the network to bioacoustic data, we modify the first convolutional layer to accept single-channel inputs and average the pretrained weights across channels. We replace the final fully connected layer with a single-output neuron for binary classification. We train the model using mini-batch stochastic optimization with a batch size of 32 over 50 epochs. We use the Adam optimizer with a fixed learning rate of approximately $2.9 \times 10^{-4}$ and fine-tune all layers jointly without freezing any part of the backbone. We do not use class weighting.

We evaluate model performance using standard binary classification metrics, including accuracy, F1-score, and area under the precision--recall curve (AUPRC). We visualize performance across folds with precision--recall curves (Fig. 6) and present results in Table 4.

\begin{figure}[!htbp]
  \centering
  \includegraphics[width=\linewidth]{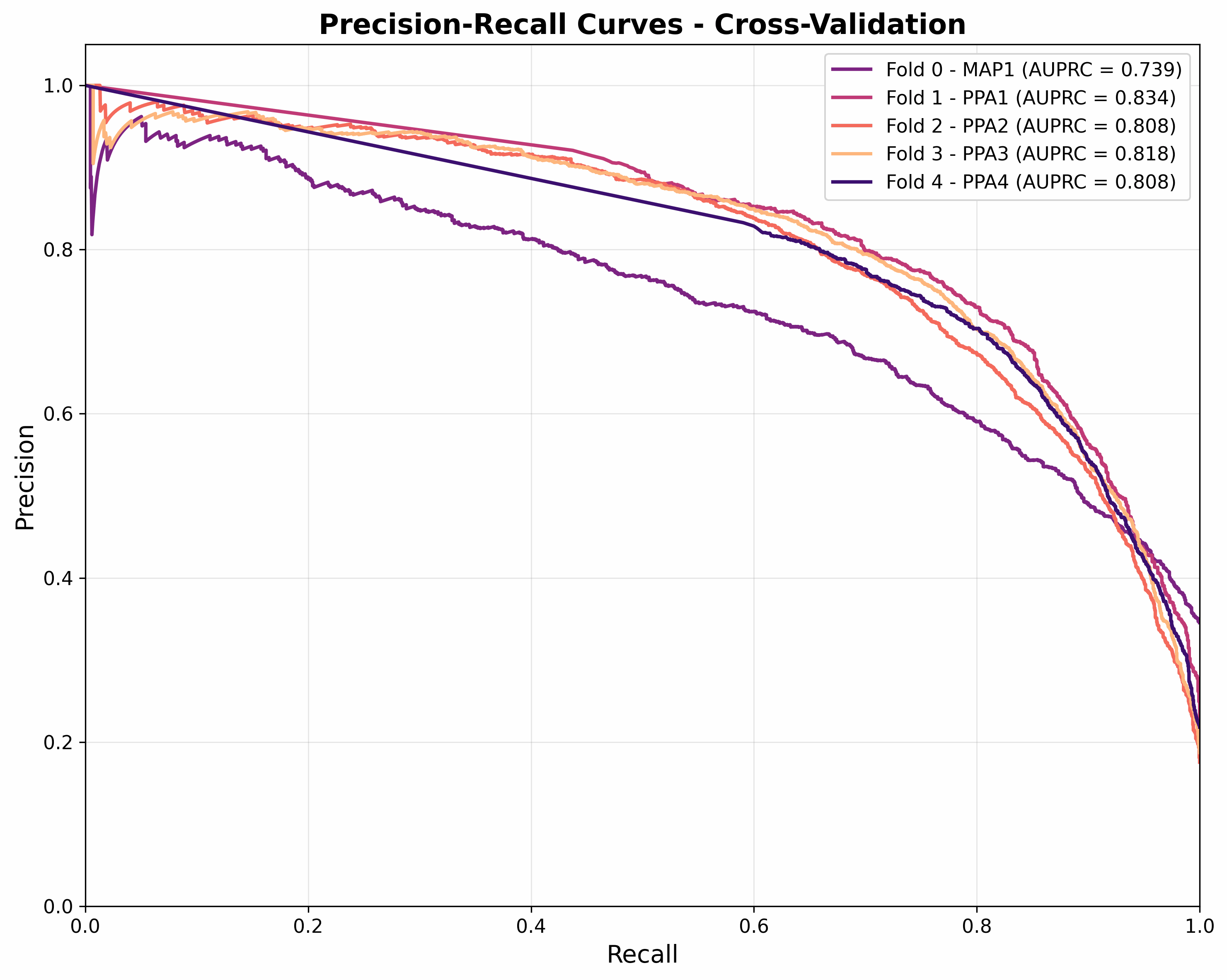}
  \caption{Precision--recall curves for the baseline bird vocalization detector, with the corresponding AUPRC shown in the legend, across the five leave-one-project-out cross-validation folds.}
  \label{fig:pr}
\end{figure}

\begin{table}[!htbp]
  \centering
  \caption{Model performance metrics from 5-fold cross-validation. The table reports F1, AUPRC, Precision, Recall, and Accuracy for each fold, along with the mean and standard deviation (Std) across all folds, summarizing the model's overall performance and variability.}
  \label{tab:cv-metrics}
  \small
  \begin{tabular}{lccccccc}
    \toprule
    \textbf{Metric} & \textbf{Fold 0} & \textbf{Fold 1} & \textbf{Fold 2} & \textbf{Fold 3} & \textbf{Fold 4} & \textbf{Mean} & \textbf{Std} \\
    \midrule
    F1        & 0.650 & 0.741 & 0.723 & 0.734 & 0.742 & 0.718 & 0.039 \\
    AUPRC     & 0.734 & 0.821 & 0.780 & 0.793 & 0.859 & 0.797 & 0.047 \\
    Precision & 0.530 & 0.794 & 0.647 & 0.728 & 0.861 & 0.712 & 0.129 \\
    Recall    & 0.840 & 0.694 & 0.818 & 0.740 & 0.652 & 0.749 & 0.080 \\
    Accuracy  & 0.688 & 0.893 & 0.890 & 0.899 & 0.880 & 0.850 & 0.091 \\
    \bottomrule
  \end{tabular}
\end{table}

Across the five folds, the baseline achieved a mean F1-score of 0.718 $\pm$ 0.039, AUPRC of 0.797 $\pm$ 0.046, and accuracy of 0.850 $\pm$ 0.091, indicating effective discrimination between bird and non-bird acoustic events across diverse recording conditions. Precision (0.712 $\pm$ 0.129) and recall (0.749 $\pm$ 0.080) are reasonably balanced, suggesting a favorable trade-off between false positives and false negatives. Per-fold performance varies more than the overall averages suggest, with Fold 0 (MAP1) the weakest and Fold 4 (PPA4) the strongest. This trend is also reflected in Fig. 6, where Fold 4 shows the highest precision--recall curve and Fold 0 the lowest. The degradation on MAP1 likely reflects ecological and acoustic dissimilarities between the Caribbean lowlands of Magdalena and the Andean--Amazonian sites in Putumayo.

\begin{figure}[!htbp]
  \centering
  \includegraphics[width=\linewidth]{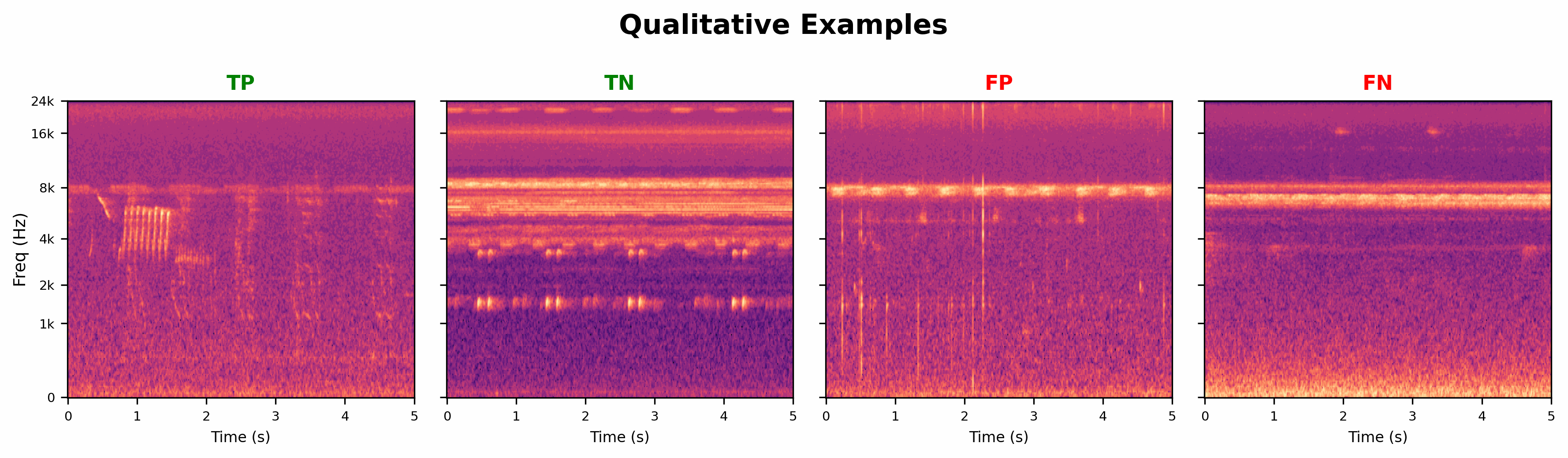}
  \caption{Qualitative examples of model predictions on the PPA4 test set (Fold 4) illustrate representative true positive (TP), true negative (TN), false positive (FP), and false negative (FN) cases.}
  \label{fig:qualitative}
\end{figure}

To further illustrate the quantitative results, Fig. 7 presents qualitative examples of model predictions for Fold 4. The true positive example shows a clear broadband bird vocalization between 1--8 kHz with a distinct harmonic structure, which the model correctly identifies. The true negative example, dominated by low-frequency ambient noise and insect-like tonal patterns below 2 kHz, is correctly classified as containing no bird activity. The false positive case shows continuous tonal energy bands around 4--8 kHz that may resemble bird vocalizations in spectral shape, likely misleading the classifier. The false negative example shows faint vocalizations that the model fails to detect, suggesting that low-frequency calls remain challenging for the current architecture. These examples illustrate that detection is comparatively easier for clear, high-SNR vocalizations but may struggle with subtle signals or ambiguous non-biological sounds that share spectral characteristics with bird calls.

Together, the quantitative and qualitative results establish PteroSet as both a rigorous training resource and a challenging benchmark for bird detection. The dataset enables model development while highlighting key challenges of real-world bioacoustic monitoring in Neotropical soundscapes, including cross-site domain shift, faint or low-frequency vocalizations, confounding non-biological sounds, dense acoustic overlap, and long-tailed species distributions. These challenges are often underrepresented in cleaner, temperate datasets but are critical for real-world deployment. By centering these conditions, PteroSet promotes the development of models that generalize beyond controlled environments and better reflect the demands of automated bioacoustic monitoring across ecosystems worldwide.

\section*{Usage Notes}
\subsection*{Discontinuous Recording Structure}

Users should note that PteroSet audio files are not continuous recordings. Each file consists of 10-second clips taken every 30 minutes over a 24-hour period (up to 48 segments per recording), so consecutive clips in a file are separated by about 30 minutes. As a result, models must not treat the entire file as continuous audio. Our baseline model avoids this by excluding any spectrogram window that spans segment boundaries, and we recommend similar precautions for other processing pipelines to prevent false temporal associations. Additionally, PPA1 recordings have overlapping segment boundaries due to a 1-second crossfade applied during compilation; the segment stride for PPA1 is 9 seconds rather than 10. Users processing PPA1 files should use the appropriate stride when identifying segment boundaries.

\subsection*{Labeling Strategy and Overlap Thresholds}

The baseline model in this study assigns a positive label to any 5-second spectrogram window that overlaps with at least one annotation, regardless of the extent of overlap. While this approach maximizes sensitivity, it can introduce label noise when only a very small fraction of the window contains annotated signal. Users may benefit from implementing a minimum overlap threshold, requiring that a specified proportion of the window duration or a minimum number of seconds contain annotated signal before assigning a positive label. The annotation duration statistics in the Data Overview section can serve as a reference for calibrating such thresholds. Exploring threshold-based labeling strategies and soft-label approaches may improve model robustness.

\subsection*{Domain Shift Between Recording Sites}

PteroSet includes recordings from two distinct regions in Colombia. These areas differ in species, habitat, noise levels, and acoustic profiles, leading to a domain shift evident in model performance. For applications across regions or at new sites, users should consider domain adaptation, site-specific fine-tuning, or balanced sampling to mitigate geographic bias in training data.

\subsection*{Class Imbalance and Species-Level Annotations}

As described in the data overview, variation in annotation strategy across projects plays a central role in shaping the dataset's composition and imbalance. Users performing multi-class classification should account for this imbalance, as standard training procedures may produce biased models. Approaches such as few-shot learning, class-weighted loss functions, or oversampling may be necessary to improve performance on underrepresented species.

\subsection*{Overlapping Annotations and Multi-Label Considerations}

Acoustic co-occurrence is prevalent in this dataset, reflecting the characteristic complexity of Neotropical soundscapes. Approximately 19.8\% of all annotations are involved in at least one temporal overlap, and among overlapping pairs, 60\% also overlap in the frequency domain. This means that a substantial proportion of spectrogram windows will contain vocalizations from multiple species simultaneously, as shown in the technical validation section. Users developing single-label classifiers or models that assume isolated acoustic events should be aware that this assumption does not hold for a significant fraction of the data. Multi-label classification formulations, in which each window can be assigned to multiple species simultaneously, are more appropriate for this acoustic environment and are likely to yield more ecologically meaningful predictions. When using the dataset for binary bird detection, as in the baseline model, overlapping annotations do not pose a labeling conflict, since any bird presence triggers a positive label. However, for species-level tasks, the co-occurrence structure must be explicitly accounted for in the labeling scheme and the evaluation protocol.

\subsection*{Extending and Integrating the Annotation Format}

One goal of this work is to promote interoperability across bioacoustic datasets by adopting the COCO-inspired annotation scheme. Researchers who maintain acoustic datasets are encouraged to convert their annotations to the JSON format described in Table 3. When merging PteroSet with external datasets, category identifiers are automatically reconciled across sources to avoid taxonomic conflicts, and sound file identifiers remain globally unique. To support this process, our repository also provides practical examples of how to combine PteroSet with other datasets using this unified schema. Consistent use of a shared format across the bioacoustics community would substantially lower the barrier to building large-scale, multi-site training corpora and facilitate reproducible benchmarking of detection and classification methods.

\section*{Data Availability}

The raw and processed datasets supporting the findings of this study are publicly available in the PteroSet Zenodo repository at \url{https://zenodo.org/records/19137071}.

\section*{Code Availability}

All code used for data processing and technical validation is available on GitHub at \url{https://github.com/microsoft/PteroSet} to ensure reproducibility.

\section*{Funding}

This work was supported by funding from Parex Resources and Amerisur Resources. These organizations provided financial support for data collection. The funders had no role in study design, data interpretation, or the decision to publish these results.

% References — thebibliography environment auto-renders the "References" heading.
% Reformat bibitem labels as "N." (Nature/Scientific Data style) instead of "[N]".
\makeatletter
\renewcommand{\@biblabel}[1]{#1.}
\makeatother

\end{document}